\documentclass[runningheads]{llncs}
\usepackage[utf8]{inputenc}
\usepackage{graphicx}
\usepackage [english]{babel}
\usepackage{xurl}

\usepackage[numbers,sort&compress]{natbib}

\begin{document}
\renewcommand{\bibname}{References} 
\title{Number and quality of diagrams in scholarly publications is associated with number of citations}
\titlerunning{Number and quality of diagrams is associated with number of citations}
\author{Guy Clarke Marshall\inst{1}, Caroline Jay\inst{1}, and Andr\'e Freitas\inst{1,2}}
\authorrunning{GC Marshall et al.}
\institute{Department of Computer Science, University of Manchester, Manchester, UK \\ \email{guy.marshall@postgrad.manchester.ac.uk, \\ \{caroline.jay, andre.freitas\}@manchester.ac.uk}
\and 
Idiap Research Institute, Rue Marconi 19, Martigny, 1920, Switzerland\\
}
\maketitle

\begin{abstract}
    Diagrams are often used in scholarly communication. We analyse a corpus of diagrams found in scholarly computational linguistics conference proceedings (ACL 2017), and find inclusion of a system diagram to be correlated with higher numbers of citations after 3 years. Inclusion of $>3$ diagrams in this 8-page limit conference was found to correlate with a lower citation count. Focusing on neural network system diagrams, we find a correlation between highly cited papers and ``good diagramming practice" quantified by level of compliance with a set of diagramming guidelines. Two diagram classification types (one visually based, one mental model based) were not found to correlate with number of citations, but enabled quantification of heterogeneity in those dimensions. Exploring scholarly paper-writing guides, we find diagrams to be a neglected media. This study suggests that diagrams may be a useful source of quality data for predicting citations, and that ``graphicacy" is a key skill for scholars with insufficient support at present.  
\end{abstract}



\section{Introduction}
Diagrams form a part of communications about AI systems, such as papers published at the Association of Computational Linguistics (ACL), a top natural language processing (NLP) conference. We argue that system diagrams are an important source of data about scholarly authorship practices in computer science, specifically neural networks for natural language processing, and have insufficient attention in many academic writing guides. Using Transactions of ACL 2017 as a corpus, we show that system diagrams are prevalent. We find that papers containing a system diagram are more likely to have a higher number of citations, perhaps indicating that their authors are effective science communicators, or that they write papers about systems, which are more highly cited. Further, papers containing more than two diagrams are found to be more likely to have a lower number of citations, and possible reasons for this are explored. 

Corpus analysis of diagrams is nascent, with recent analysis into connecting lines in data visualisations \citep{lechner2020modality}. We use a corpus-based approach to examine diagrams within a wider social context, and have designed our approach to leverage existing document-level citation metrics, allowing quantitative analysis. Our main contribution is to test for compliance with an existing set of neural network system diagram guidelines, using a corpus-based approach. In summary, we find system diagrams are prevalent, occurring in 82\% of papers at ACL 2017, and that diagrams in highly cited papers are more likely to contain ``good diagrams" in the sense of conforming to an existing set of guidelines. 

\section{Background}
\subsection{NLP systems}
Natural Language Processing is a discipline within Computer Science, and is concerned with creating systems that solve tasks relating to Natural Language interpretation. NLP systems take a text input, go through data manipulation steps, and create an output that is usually a classification or a prediction, such as what the next word in a sequence is likely to be. The state-of-the-art systems are technically complex, requiring application of mathematical and algorithmic techniques. These NLP systems are often described through diagrams. We have chosen to examine scholarly neural network systems, described in diagrams within NLP conference proceedings. 
\subsection{NLP system components}
Modern NLP systems are often based on neural networks, and it is these systems we focus on. A neural network takes an input (in NLP, text), and then processes this via a series of \emph{layers}, to arrive at an output (classification/prediction). Within each layer are a number of \emph{nodes} that hold information and transmit signals to nodes in other layers. Specific mathematical functions or operations are also used in these systems, such as sigmoid, concatenate, softmax, max pooling, and loss. The \emph{system architecture} describes the way in which the components are arranged. Different architectures are used for different types of activities. For example Convolutional Neural Networks (CNN), inspired by the human visual system, are commonly used for processing images. Long Short Term Memory networks (LSTM), a type of Recurrent Neural Network (RNN) which are designed for processing sequences, are often used for text.

These neural networks ``learn" a function, but have to be trained to do so. Training consists of providing inputs and expected outputs, allowing the system to develop an understanding of how an input should be interpreted. The system is then tested with unseen inputs, to see if it is able to handle these correctly. System diagrams almost always depict the training process. A more detailed introduction to LSTM architectures, including schematics, is provided by \citet{olah2015understanding}.

\section{Method}


We use ACL 2017 scholarly papers as a corpus from which to extract diagrams, because it is an appropriate size for analysis (195 long papers), is distributed with a CC4.0BY licence, and is recent enough to be relevant whilst allowing for short-term (3 year) citation analysis. 
Web of Science contains statistics of peer reviewed citations which protects to some extent from the hype surrounding various topics, and we use ``Times cited, WoS Core". Using a chi-squared test we found this metric highly correlated with the less curated ``Times cited, All"). 

Our method follows \citet{lee2017viziometrics}, adapted to use a manual extraction process in order to reduce systematic omission and make use of the validity ensuring method of \citet{lechner2020modality}. 
\begin{enumerate}
    \item Using Web of Science, publication metadata was manually extracted from all long papers from ACL 2017, including number of citations.
    \item Every figure that displayed a diagram was manually extracted, except figures in the Results section. We added diagram count as additional paper metadata. The term ``diagram'' is used to describe a conceptual diagram, usually a figure, which is not reporting results, displayed as a table, nor describing an algorithm. In practice, it encompasses system diagrams, parts of systems, graphical representations of algorithms, concept maps, flow charts of methods or systems, and formal diagrammatic languages such as parse trees. 
    \item Diagrams were stored as separate image files, labelled according to which paper they were extracted from.
   \item In each paper, at most one diagram was identified as the primary system diagram. Where multiple system diagrams were found, the one with the largest number of graphical elements was used. Additional metadata was captured for the neural network system as follows: Conformity to each individual guideline, colour or monochrome, and allocation of the diagram to one of five visual categories \citep{marshall2020understanding} and one of four mental model categories \citep{mindandmachines,rouse1986looking}. In most cases, the diagram contained features of several categories, and the most prevalent category was captured. 
   \item  Following the method of \citet{lechner2020modality}, inter-rater reliability was measured to validate scoring of guidelines compliance, on a subset of 15\% of the resulting NN system diagram (17/119). This resulted in 204 pairs of pieces of metadata scored as ``true", ``false", or ``not applicable". Using this, Gwet's $AC_1$ coefficient was calculated \citep{gwet2014handbook}, finding ``good" reliability when considering the guidelines as a set. Individual guideline conformity was variable, with Guidelines 2, 4, 7, 10 and 12 (in the ordering of Table \ref{tab:conformity}) scoring a less than ``good" Gwet's $AC_1$, and required further clarification beyond the guideline text alone to agree scoring. Subsequent assessment was done with a single coder. This manual coding resulted in the addition of over 1,600 pieces of diagram metadata, together with 400 additional paper metadata items (diagram count, and system diagram inclusion). 
   \item The conference area of each paper was manually extracted, as defined by ACL organisers \cite{aclpapersblog}.
   \item Data were analysed in R \citep{rsoftware}, using the ggplot2 package to create graphics \citep{rsoftwareggplot2}.
\end{enumerate}


\section{Results}

\subsection{Diagrams in context}

\begin{figure}
    \centering
    \includegraphics[width=0.8\textwidth]{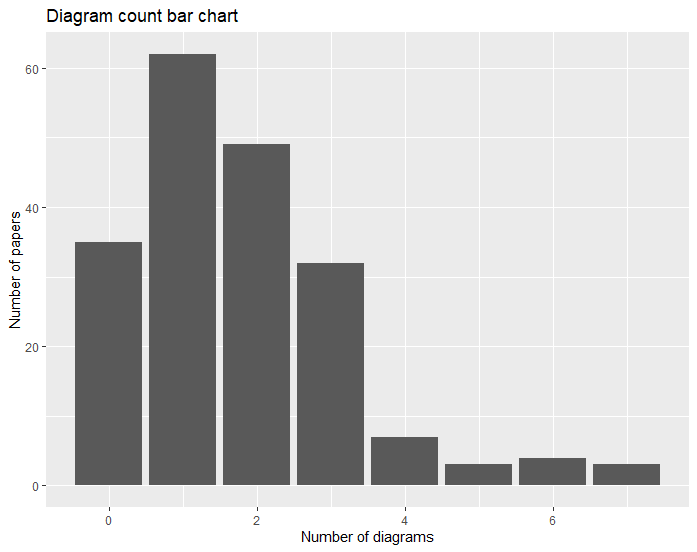}
    \caption{Number of (non-results) diagrams in ACL 2017 papers is normally distributed, with inclusion of one or two diagrams most common. Most papers include a diagram.}
    \label{fig:diagramcountbarchart}
\end{figure}

Fig.~\ref{fig:diagramcountbarchart} shows the frequency of diagrams in ACL 2017 proceedings. The large number of papers, particularly highly cited papers, which include system diagrams demonstrates the importance of system diagrams in communicating at ACL 2017. 

\begin{figure}[htbp]
    \centering
    \includegraphics[width=0.8\textwidth]{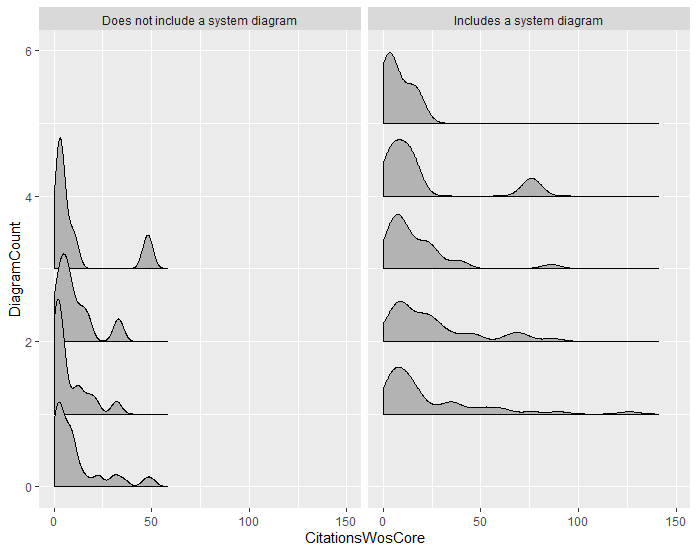}
    \caption{Paper's number of citations ridgeline plot, split by number of diagrams included in the paper, and whether at least one of the diagrams is a system diagram.}
    \label{fig:citationsystemflag}
\end{figure}

Fig.~\ref{fig:citationsystemflag} shows the distribution of citations, split by whether the paper includes a system diagram. Note the higher mean number of citations for those including a system diagram. Fig.~\ref{fig:citationsystemflag} also shows that for 1-2 diagrams, inclusion of a system diagram is correlated with a significantly higher number of citations. For these papers, a system diagram is correlated with 19 more citations (from 7.4 to 26.4 citations, over 250\% higher). 

Other factors are likely to be involved with number of citations, including author reputation, research trends, social media presence, and so on. Note that for $>$4 diagrams, the lines converge, plateauing to a low number of citations. Having more than two diagrams being correlated with a lower number of citations could be explained by: (a) the contribution itself may not be narrow enough to provide depth of contribution (b) the authors may not have put sufficient care into writing the paper(!) (c) diagram space reduced space for other important content. It is unlikely to be only domain specific reasons (such as inclusion of existing sub-domain-specific formal diagrammatic languages) as conference area was not correlated with diagram count. 

To summarise the key insights:
\begin{itemize}
\item 160/195 (82\%) of all ACL 2017 papers included diagrams to represent concepts (not including results or algorithms)
\item 124/195 (64\%) of all ACL 2017 papers included at least one system diagram. 

    \item Including 1-2 diagrams, of which at least one is a system diagram, is correlated with a 250\% higher number of citations.
        \item Having more than two diagrams is correlated with lower number of citations (see Fig.~\ref{fig:citationsystemflag}). In a linear model each additional diagram is correlated with 5.6 fewer citations (p=0.02). In the subset of papers which include a system diagram, this effect increases to 7 fewer citations per additional diagram.
        \item 82/119 (69\%) of neural network systems diagrams used colour, which may affect accessibility.
         \item Diagrams may be a valuable source of data for modeling number of citations. See Section \ref{section:predictionmodel}. 
         \end{itemize}

\subsection{Conference areas}
In an attempt to remove some of the effect of the content of paper, we analysed whether there was a relationship between the 17 conference areas \cite{aclpapersblog} and (i) citations (ii) inclusion of a system diagram (iii) number of diagrams (iv) usage of examples. We found no significant difference between pairs of these attributes using chi-squared tests, using the entire dataset.

Every conference area (except the solitary paper in the Biomedical area) included at least one system diagram. 
Together, this suggests that system diagrams are important across all areas of ACL, not limited to certain sub-domains. To further investigate any potential paper-content-related cause, we found 21 papers contain the word ``architecture", and 18 of those contain a system diagram. Number of citations and the abstract containing the word ``architecture" are correlated (p-value $<$ 0.01), with those containing ``architecture" having on average 20.4 more citations (than 15.9). As would be expected, the abstract containing the word ``architecture" and including a system diagram are not independent: There is a significant relationship (p-value $<$ 0.05). Causality is therefore ambiguous, as to whether architectural papers are more likely to be highly cited, or whether it is due to the presence of the diagrams. In either case, system diagrams are important in this corpus.




\subsection{Neural network system diagram guideline conformity}


119 of 124 system diagrams described neural network systems (the others being diagrams of an embedding only, or not a neural system). These 119 diagrams were assessed against each of the 12 guidelines established following an interview study \citep{marshall2020researchers}, which are reproduced as part of Table \ref{tab:conformity}. These guidelines were chosen in favour of other diagramming guidelines due to their domain specificity.  
\begin{figure}
    \centering
    \includegraphics[width=0.8\textwidth]{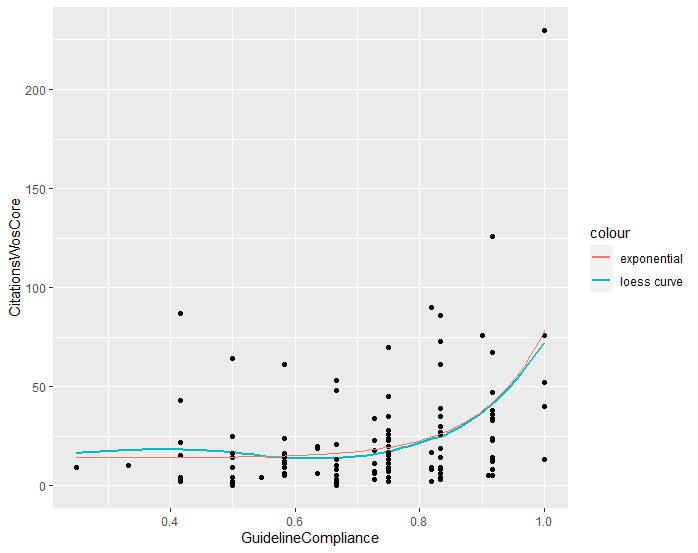} 
    \caption{Scatter plot of number of citations versus NN system diagram guideline compliance (as a quantitative proxy for ``how good the diagram is"). LOESS curve for locally weighted smoothing is in blue, and the function $y=e^{10(x-7/12)} + 14$ is in red.}
    \label{fig:guidelinecompliance}
\end{figure}

Whilst the analysis was exploratory, our main hypothesis was that we would observe fewer diagram guideline violations in papers with a higher number of citations. In this analysis, we found a correlation between number of citations and ``specific" (p$<$0.05), and also ``self contained" (p$<$0.05) guidelines. The other guidelines alone did not correlate with a significant difference in number of citations. However, the best correlation was found with an average of the guideline compliance. The results of this are shown in Fig.~\ref{fig:guidelinecompliance}. Papers with diagrams conforming with 10/12 guidelines or greater are likely to have a higher number of citations than those conforming to fewer than 10 guidelines. 




The LOESS curve in Fig.~\ref{fig:guidelinecompliance} can be approximated by an exponential function, $citations=e^{10(compliance-7/12)} + 14$, where ``$7/12$" captures the increase in citations observed from higher levels of compliance, ``$14$" captures the asymptotic average number of citations for low compliance papers, and the multiplier ``$10$" fits the curve. The only independent variable is average guideline compliance in each diagram, and we do not aim to model citations accurately, but only to argue that the guidelines capture some diagramming behaviours of effective communicators. 









\begin{figure}
    \centering
    \includegraphics[width=0.7\textwidth]{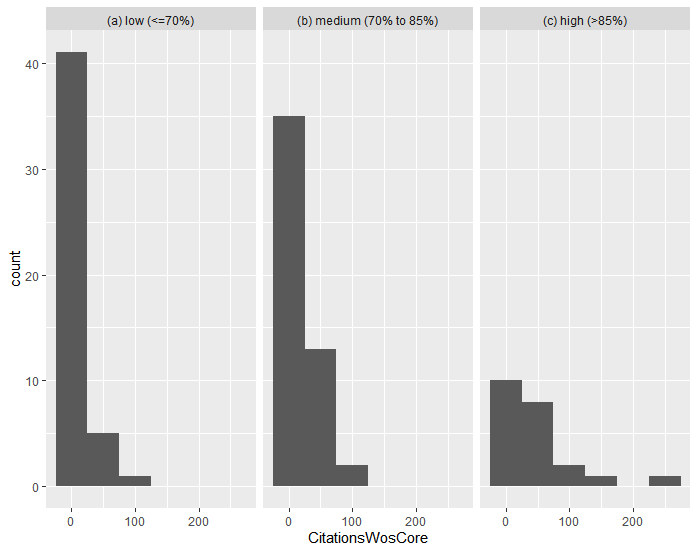}
    \caption{Number of citations for papers containing system diagrams, grouped by level of guideline compliance.}
    \label{fig:compliancehistogram}
\end{figure}

For diagrams conforming to fewer than eight guidelines, it appears visually almost to be random whether the author has conformed to the guidelines or not. Fig.~\ref{fig:compliancehistogram} shows diagrams with high, medium and low compliance (the percentages were chosen to make the population sizes similar). This simplifies the results of Fig.~\ref{fig:guidelinecompliance}, and shows that papers containing diagrams complying with over 85\% of guidelines are less likely to have a lower number of citations, and more likely to have a higher number of citations.

Of diagrams conforming to 11/12 guidelines, six used more than one arrow, two did not use examples, one used unconventional objects, and one violated the colour guideline. The reason for the multiple arrow types in each case was often evident by inspection, usually to separate a type of data flow or distinguish between abstraction levels (e.g. mathematical workings of a neuron vs data pipeline). It may be that the ``one type of arrow" guideline should be revisited through user evaluation. The authors are evidently good communicators, in terms of having above-average number of citations, and this guideline not being observed by those authors supports the view that neural network system diagrams may be better supported by flexible guidelines rather than rigid standards \citep{marshall2020researchers}.

In an attempt to identify whether diagrams were formed using good diagramming practices, we examine conformity of the guidelines in the top quartile of citated papers including a NN system diagram (of the 119 NN system diagram papers, this includes the 30 papers with 28 or more citations). ``Not applicable" scores were omitted from the analysis. Table~\ref{tab:conformity} shows the results. No top quartile paper violated the self-contained guideline, meaning those diagrams were understandable on their own and did not reference text directly. Including input and output, which is often related to self-containedness, was also done by all except three of the top 27 papers. Further, whilst avoiding unconventional objects was done in 25/30 top quartile diagrams, this perhaps came at some cost of using precision with care, where that is conventional (12/30 not conforming to this). Almost half (14/30) of top quartile papers used multiple arrow types, suggesting that thoughtful use of multiple arrows may be good for communicating different abstraction levels. Using an example in the diagram was done in 19/30 highly cited papers, and those that did not include an instantiated example often included mathematical notation (11/19). We maintain that having an example is useful, and mathematical equations also facilitate this process.




\begin{table}[htbp]

\begin{tabular}{l|rrrr}
\hline
Number of citations quartile       & Bottom (30 papers)  & 2nd (31)  & 3rd (28) & Top (30)  \\ \hline
No prevalent unconventional objects & 3  & 3  & 2  & 5  \\
One arrow for information           & 17 & 17 & 17 & 14 \\
Precision care                      & 15 & 11 & 8  & 15 \\
Input and output                    & 2  & 4  & 4  & 4  \\
Example                             & 18 & 16 & 10 & 11 \\
Meaningful visual encoding          & 10 & 7  & 8  & 3  \\
Easy navigation                     & 7  & 8  & 7  & 4  \\
Colours not aesthetic only          & 12 & 13 & 16 & 8  \\
Conventions                         & 12 & 8  & 7  & 4  \\
Expectation matching                & 6  & 5  & 7  & 1  \\
Specific                            & 17 & 11 & 15 & 5  \\
Self contained                      & 6  & 6  & 5  & 0 \\ \hline
\end{tabular}
\caption{Guideline violation count, for each quartile of number of citations}
\label{tab:conformity}
\end{table}



Of the 119 NN system diagrams, 64 diagrams (54\%) contained an explicit example to instantiate the input. If an example was used, it was certainly used in the input, and often in the output, and occasionally at intermediate steps. There was not a significant difference in citations whether the paper's system diagram included an example or not, and both categories appear to follow a similar distribution, though the most highly cited papers included an instantiated example.


Of the 30 top-cited papers, nine did not use colour, 13 used colour meaningfully, and eight used colour for aesthetics only. This suggests that colour for aesthetics may be appropriate in some cases (or it may be emphasising aspects of the system in a manner not uncovered by our method). However, examining relative frequency of occurrence, we find that more highly cited papers are less likely to use colours for aesthetics only.

Guideline compliance was not correlated with abstract including an ``architecture" keyword, nor with conference area, suggesting that these guidelines are equally applicable to each particular contribution type.



\subsection{Classification of neural network system diagrams}
To further investigate the diagram author choices, we classify each neural network system diagram into one of the following categories: Visual (two types: circles or blocks), Math (essentially a diagrammatic description of an algorithm), Lightweight (e.g. block diagram with very few visual components), or Unorthodox (unusual diagrams). These categories are based on \citet{marshall2021structuralist}, in which semiotic classification of diagrams was conducted on a random sample of 40 diagrams from corpora of natural language processing and neural network conference proceedings (COLING 2016 \& 2018, NIPS 2016 \& 2017, NAACL 2016 \& 2018, EMNLP 2016 \& 2017). These venues are similar in scope to ACL. We chose to use this external categorisation to avoid over-fitting to the present corpus, which appeared to be a risk due to the heterogeneity and large number of possible ways to classify the diagrams. The categories proposed as a result of this previous analysis also appear relevant and provide reasonable coverage of the ACL 2017 corpus: Using these categories, we identified 58 Visual (rectangle), 19 Visual (disk), 17 Lightweight, 13 Mathematical and 12 Unorthodox diagram types.

We found no significant difference with citations using these groups. Each diagram type has a normal distribution of number of citations with a similar mean, and chi-squared tests found no significant differences. 

\begin{table}[htbp]
\label{table:classifications}
\begin{center}
\begin{tabular}{l|lllll}
\hline
Classification  & Block  & Disk  & Lightweight & Mathematical & Unconventional  \\ \hline
Form     & 17 & 3  & 7 & 4 & 3  \\
Function & 8  & 3  & 6 & 4 & 2 \\
Purpose  & 2  & 1  & 4 & 0 & 1 \\
State    & 31 & 12 & 0 & 4 & 6  \\ \hline
\end{tabular}
\end{center}
\caption{Prioritised visual and mental classifications for each diagram, showing coverage of most combinations}
\end{table}

Another NN system diagram categorisation \citep{mindandmachines}, based on whether Form, Function, Purpose or State was preferentially displayed in the diagram also did not produce any significant results, even when combined with other cognitively-based data such as example inclusion or vector visualisation types. Table \ref{table:classifications} shows the results, which reflect the broad range of priorities in terms of framing content, not just visualisation.

\section{Related work}

\subsection{Scholarly figure analysis}
Much attention is given to the automated extraction of information from scholarly figures, including the classification of charts into bar charts, pie charts, etc. \citet{roy2020diag2graph} recently created a classification system for neural network system diagrams. Their system classifies deep learning architectures into six categories, (e.g. 2D boxes, pipeline) based on how the layers are visually represented. This, and many other scholarly processing systems, rely on pdffigures 2.0 \citep{clark2016pdffigures} for diagram extraction, which has known limitations and edge-cases. In particular, some types of figure (such as those with an L-shape) are systematically omitted \citep{pdffiguresreadme}.

\citet{marshall2020understanding} apply the Richards-Engelhardt framework, recently updated as VisDNA \citep{engelhardt2020dna}, to neural network system diagrams. The complexity and internal inconsistency of the diagrams was found to make application of VisDNA challenging. \citet{marshall2020researchers} conducted an interview study on the role of diagrams in scholarly AI papers, which reported 12/12 participants using diagrams to get a summary of the paper, and found some participants (3/12) used the diagram before any text in the paper. 

Medical scholarly publications are a common domain for automated figure analysis, particularly using PubMed Central, a medical bibliographic database. PubMed contains metadata, such as figure captions, and often itemised figure files, which are useful for automated processing. \citet{lee2017viziometrics} created a visual PubMed scholarly diagram search engine, which clusters diagrams based on visual features (into high level categories e.g. diagram, graph, equation). In their accompanying analysis of figure usage patterns over time and with relation to impact, they found that ``higher impact papers tend to have higher density of diagrams and plots." They term their analysis, software and wider research agenda ``viziometrics." 

\subsection{Academic figure-writing}
\citet{carberry2006information} note that information sometimes resides in figures that cannot be found elsewhere in the text. This suggests that diagrams contain content not available elsewhere, and as such may have an important and unique role when reading and extracting information from a paper. 



Discussion of figures and diagrams is scarce in popular academic writing guides. Swales and Feak's ``Academic Writing for Graduate Students" \citep{swales2004academic}, despite including 11 conceptual diagrams to explain their own work, only gives guidance for the use of charts, not for the use of other figures such as system or conceptual diagrams. The only support for diagramming provided by \citet{murray2009writing}, across 212 pages, is the prompting question ``Do you have any figures, diagrams or tables to include?".
Schimel's ``Writing Science" \citep{schimel2012writing} includes limited advice on referencing a chart in the text, and their advice on diagrams and figures extends only to the following comment: ``I have always felt that I don't understand something until I can draw a cartoon to explain it. A simple diagram or model - the clearer the picture, the better". A lack of support for ``graphicacy" (as a skill alongside numeracy, literacy and articulacy) \citep{balchin1972graphicacy} has also been identified in science textbooks \citep{betrancourt2012graphicacy}.


There are exceptions to this brevity. One such guide providing deeper diagramming advice is ``Writing for Computer Science" by \citet{zobel2004writing}, which includes one dedicated chapter and two additional subsections elsewhere covering topics relating to figures. The discussion includes tables, algorithm figures, graphs, and use of figures in slide presentations. \citeauthor{zobel2004writing} notes that ``Diagrams illustrating system structure often seem to be poor. In too many of these pictures the symbolism is inconsistent: boxes have different meanings in different places, lines represent both control flow and data flow, objects of primary interest are not distinguished from minor components, and so on." 

\subsection{Number of citations as a scholarly metric}
Number of citations is an important performance indicator for researchers \citep{aksnes2019citations}, and is widely available on scholarly information platforms. There are numerous well known issues with citations, including self-citation \citep{baird1994citations} and bias towards positive results \citep{duyx2017scientific}. However, \citet{radicchi2017quantifying} conclude that ``when scientists have full information and are making unbiased choices, expert opinion on impact is congruent with citation numbers." 
Predicting citations is an important part of Scientometrics (e.g. \citet{bai2019predicting}), the study of quantitative aspects of science. Along with content of the paper and author properties, social factors such as Twitter \citep{luc2020does} and centrality in knowledge networks based on keywords \citep{guan2017impact} are used in recent models. 

It is common to consider long-term number of citations, usually 10 years. The most successful models predicting citations utilise past author influence, productivity, sociability, and venue features \citep{yan2012better}, obtaining an R-squared of 0.836. They do not include figure usage. They found that ``Unexpectedly, paper content is proved to have the least significance". 

With the above related work supporting its usage, and with ready availability of citations, we use number of citations as a metric in our study. As noted by \citet{leydesdorff2016citations}, ``Citation impact studies focus on short-term citation, and therefore tend to measure not epistemic quality, but involvement in current discourses in which contributions are positioned by referencing." Following this observation, we stress that in our use of citations there is no claim about the quality of the paper. 

\section{Discussion}

\subsection{Limitations}
\begin{itemize}
    \item Our corpus analysis is based on one year of one venue, and cannot be generalised. 
    \item The manual data extraction process does not scale well.
    \item Number of citations can be affected by many other factors, including author institution, author name, twitter presence, and so on. We mitigate venue by restricting to one venue. We do not take action to reduce the impact of other factors, focusing analysis on features of diagrams.
    \item Our inter-rater reliability covered only guideline conformity, not the diagram extraction or classification of figures.
    \item Whilst using the guidelines alone provided ``good" inter-rater reliability, raters needed to make subjective judgements, and required more than the guidelines alone to ensure replicability.
    \item Unlike \citet{lee2017viziometrics}, we examine only diagrams, not all figures. 

\end{itemize}

\subsection{Opportunity for utilising diagrams in models to predict number of citations}
\label{section:predictionmodel}

A simple linear model for number of citations in the ACL 2017 corpus can be made using only (i) whether the abstract contains the word ``architecture'' and (ii) the level of conformity of the system diagram to a set of guidelines (zero if a system diagram is absent). This model has an R-squared of 0.132, suggesting that 13.2\% of the variation in number of citations can be explained by these two factors alone. This simple model performs comparatively to existing citation predictions based on the entire text of the paper \citep{yan2012better}, which report 0.13 R-squared on a different scholarly corpus. Richer state-of-the-art models using social variables have R-squared around 0.4 at the 3rd year, again on a different scholarly corpus \cite{abrishami2019predicting}. This supports the central claim of the utility of diagrams in the medical-centric ``viziometrics" research agenda of \citet{lee2017viziometrics} and suggests figures may be a underutilised data source more broadly for scientometrics.


\subsection{Diagrams as an unexplored data source for author communication}

As noted in the related work, diagrams are regularly omitted from analysis of scholarly documents, including those in Scientometrics. Our finding that properties of system diagrams are correlated with number of citations suggests they warrant more consideration. This work provides further motivation for improved scientific scholarly graphicacy, the benefits of which to are often pedagogically focused. ``Drawing to learn" is an active research area \citep{ainsworth2021learning}, and there have been studies on benefits of drawing for scientific thinking specifically (see \citet{fan2015drawing}). Our findings support the centrality of diagrams in scholarly communication previously identified in Medical Science \citep{yang2019identifying}, and lends weight to the reported primacy for some users of diagrams within the AI scholarly context \citep{marshall2020researchers}.

\section{Conclusion}
Diagrams are an important, prevalent, and neglected component of scholarly communication about neural network systems, and diagramming is not proportionately discussed in many scholarly writing guides. At ACL 2017, high quantities of diagrams were found to be correlated with lower numbers of citations. The inclusion of system diagrams was found to be correlated with higher numbers of citations, suggesting their usage is a good scholarly communication practice in this domain.

We have shown good domain-specific diagramming practices, quantified by compliance with a set of guidelines, to be correlated with a higher number of citations for ACL 2017 papers. Two diagram classification schemes, based on mental models and on visual encoding, were not found to be correlated to number of citations, instead demonstrating the heterogeneity of the design space. This research has shown that diagrams are important for communicating about scholarly neural network systems, and may be an underutilised tool for understanding and improving scholarly communication.


\appendix

\renewcommand{\bibsection}{\section*{References}}
\bibliographystyle{splncs04nat}
\bibliography{bib}
\end{document}